\documentclass[twocolumn,showpacs,preprintnumbers,
amsmath,amssymb,aps,prd,nofootinbib,superscriptaddress,
eqsecnum]{revtex4}
\usepackage{graphicx}
\usepackage{dcolumn}
\makeatletter
\renewcommand{\p@subsection}{}
\makeatother

\newcommand{\Rmnum}[1]{\expandafter\@slowromancap\romannumeral #1@}

\newcommand{\be}{\begin{eqnarray}}
\newcommand{\ee}{\end{eqnarray}}

\def\lsim{\mathrel{\rlap{\lower3pt\hbox{\hskip1pt$\sim$}}
     \raise1pt\hbox{$<$}}} 
\def\gsim{\mathrel{\rlap{\lower3pt\hbox{\hskip1pt$\sim$}}
     \raise1pt\hbox{$>$}}} 

\def\la{\langle}
\def\ra{\rangle}
\def\bi{\bibitem}

\begin{document}

\title{The ``Folk Theorem" on Effective Field Theory:\\  A Case for Nuclear Physics }

\author{Mannque Rho}

\affiliation{%
Institut de Physique Th\'eorique, Universit\'e Paris-Saclay, CNRS, 
CEA, 91191 Gif-sur-Yvette c\'edex, France }

\begin{abstract}
Although largely unrecognized,  Gerry Brown had played a seminal and prescient role for the development of the currently heralded ``first-principles approach" to nuclear dynamics known as ``nuclear effective field theory" ($n$EFT for short). I give a brief account in what way he entered -- together with his Korean colleagues  -- at the earliest stage of its development and what he left behind for the future in paving the road to go  {\it way beyond} the standard $n$EFT. I do this in the context of Gerry's original conceptual ideas  conceived after the ``dilepton fiasco" that indicate a surprising new role of hidden local and scale symmetries in nuclear physics.
\end{abstract}
\date{\today}
\maketitle
In the Spring 1981, Gerry Brown and I were invited to lecture at the Erice School organized by Dennys Wilkinson on ``Quarks and the Nucleus"~\cite{wilkinson}. Gerry lectured on ``Nucleon-nucleon forces, quarks and boson exchange" and my lectures were on ``Pions and the chiral bag." Approximately delineated, Gerry was focused on interactions dealing with quarks and heavy bosons,  i.e.,  ``hard" degrees of freedom, and I was on the pions, hence  ``soft" interactions.  This separation was made in terms of the ``chiral bag" with the MIT bag in which asymptotically free quarks are confined, clouded outside by  pions. What motivated us in our lectures was to arrive at the picture of, say, the $^{208}$Pb nucleus described with 208  ``Little Bags" clouded by pions replacing the big MIT bag with confined 624 quarks, considered at the time more realistic by QCD for nuclear physics.  The former  taking the ``little bag" point-like would be the picture closest to what had been adopted in nuclear theory until then. The power of the little chiral bag was that it provided the possibility to take into account the very short-distance physics encoded in QCD not feasible with meson exchanges, thereby reconciling with the asymptotically free QCD dynamics. 

It was at the Erice School that we discussed (to the best I know, the first time in nuclear physics) how to treat the pion dynamics systematically to account for the nucleon structure in nuclei by drawing the inspiration from Weinberg's chiral dynamics,  going beyond the phenomenological Lagrangian that encodes soft-pion theorems~\cite{schwinger-fest}. Gerry left the task of going beyond soft-pion theorems to me while he was concerned with the heavier degrees of freedom, namely, the vector mesons $V\equiv (\rho,\omega)$ and perturbative quarks in nuclear physics. My goal there was  to compute higher chiral-order terms in the pion sector that would lead to excitations that are heavier than soft pions and match to the mass scale corresponding to that of the vector mesons $V$ and the scalar meson $\sigma$. This meant that the power counting should consist of going, put in the present-day jargon in the field, to N$^n$LO for $n>2$.
Since the precise {\it direct} confrontation with Nature was much more feasible with electroweak (RW) response functions, my role was to mainly focus on computing higher-chiral order corrections to nuclear magnetic moments and axial-vector matrix elements. Gerry was not directly involved in this direction, but he followed what transpired in the development between the Erice school and early 1990's which turned out to be spectacularly successful as I will mention below.  He was convinced the success met in EW response functions would apply as well to nuclear forces since nuclear forces and EW response functions are intimately connected by the Ward-Takahashi identities in nuclei. For this reason he did not engage directly in the specific working-out of the chiral expansion in nuclear forces involving pions. What is perhaps more significant was the revival of the skyrmion model for nucleons in the context of QCD in 1982 and the formulation  of the Cheshire Cat Principle (CCP) in 1985~\cite{CCP}  which took away his and his collaborators' attention from nuclear chiral perturbation theory to the impact of soliton structure in the nuclear dynamics. Most significantly the Cheshire Cat Principle allowed at low energy the MIT bag to be ``shrunk" to zero size without changing physics with the pion cloud taking over nuclear dynamics completely, {\it provided the ``smile" of the Cheshire Cat remains in the cloud.} This meant that the short-distance physics embodied by asymptotic quark interactions be captured {\it mostly}, if not entirely, by the ball of pions, generating the heavy  degrees of freedom (DoFs).

Before Weinberg's celebrated first paper on nuclear forces from chiral Lagrangian appeared at the beginning of 1990's~\cite{weinberg1,weinberg2}, there had been an exchange of discussions on nuclear forces between Weinberg and Gerry. Gerry faxed to me a copy of Weinberg's draft of \cite{weinberg2} while I was in Seoul. This paper made me extremely excited because it corrected a mistake in the counting rule that I had made in the 1981 Erice lecture. This correction together with the  basic strategy of Weinberg's counting rule (called ``Weinberg counting" in the literature) made me finally see why the soft-pion corrections in the nuclear EW response functions turned out to work so well in some of what were considered as ``fundamental" nuclear processes. In particular, it explained surprisingly simply why there were only a few percent corrections coming from two-body currents (called ``meson-exchange" currents then) to, among others, such classic processes in nuclear physics as the thermal neutron-proton radiative capture and the proton-proton solar neutrino process, whereas soft pion-exchange two-body terms coming at NLO made a whopping 100\% correction to the axial charge (first forbidden) matrix elements. This phenomenon was dubbed, in the absence of a better terminology, ``chiral filter phenomenon." I quickly drafted an article for Phys. Rev. Lett.~\cite{MR91} and showed it (together with other supporting results) to Gerry and convinced him that what we were discussing in the 1981 Erice Lectures was ``on the right track" in consistency with Weinberg's approach.  Weinberg acknowledged this in his 1991 paper~\cite{weinberg2} as well as in \cite{FT}.   

Gerry fully agreed with Weinberg for the ``soft" component involving pions  but was in disagreement on the absence in Weinberg's treatment of heavy meson fields (called H for short) for the ``hard" component, i.e., the vector mesons $V$ and the scalar meson $\sigma$  (or $f_0(500)$).   Gerry's point was that although low-energy excitations of {\it certain} H-meson channels could well be simulated at normal nuclear matter density by calculable ``irreducible" loop-order terms, certain qualitatively important terms could be largely, if not entirely, missed at higher orders required to go beyond the soft scales as relevant to dense nuclear dynamics. Examples cited, among others,  were processes quite sensitive to the nuclear tensor forces that involve interplay between ``soft" and ``hard" degrees of freedom, certain EW processes where the vector dominance (VD) plays a key role such as in dilepton properties in heavy-ion collisions, and specially massive compact stars, in all of which Gerry was keenly interested then.   His main point was that the chiral perturbation in $n$EFT anchored on soft-ness of pionic interactions could become more or less ineffective when the energy scale is of the {\it effective} in-medium vector-meson mass which can be lower than the free-space mass in nuclear medium. Given that the effective mass of the H mesons is expected to drop in nuclear medium as the vacuum is changed by increasing density or temperature, the usual chiral perturbation expansion with pions only would  break down unless the vector mesons figure explicitly.  This led to incorporating the vector-meson DoFs -- together with the scalar dilaton of comparable mass -- in hidden local symmetry Lagrangian implemented with scale symmetry in extending the DoFs from pionic to higher-mass scales.   A paper addressing this issue~\cite{BR91} appeared almost simultaneously with Weinberg's paper~\cite{weinberg2}. This paper presaged the way for Gerry to go {\it beyond} the standard $n$EFT.

To see in which way Gerry envisaged the step forward, the most appropriate is what's in  Weinberg's ``Folk Theorem" on effective field theory~\cite{FT}\footnote{Weinberg attributes this title for the theorem to Arthur Wightman.} as applied to nuclear physics. The ``Folk Theorem" stipulates:  Given that nobody knows what the putative ``Theory of Everything" -- assuming it exists --  is,  the best one can do is do an effective field theory (EFT): To quote, {\it when  you use quantum field theory to study low-energy phenomena, you are not making any assumption that could be wrong, unless of course Lorentz invariance or quantum mechanics or cluster decomposition is wrong, provided you don't say specifically what the Lagrangian is. As long as you let it be the most general possible Lagrangian consistent with the symmetries of the theory, you're simply writing down the most general theory you could possibly write down.} There can obviously be no ``rigorous proof" to this theorem, so the quotation mark on ``Folk Theorem (FT)."  The only thing one can say is how can such a theory go  wrong? 

In short, the motto for the ``FT" is: {\it Be on the right track, work harder and  of course, without making errors!}

One of the most indispensable ingredients for EFT is the energy/length scale involved. For different scales, different degrees of freedom become relevant for different low-energy limits of QCD, each with its characteristics, scales and ranges of applicability~\cite{pionless}. Thus when one is dealing nuclear processes where the energy scale is much less than the pion mass, then one can just integrate it out without violating basic ingredients and work with the ``pionless EFT" with nucleons only.  And it works very well over a wide range of problems at very low energy~\cite{pionless}.  When the scale involved is above $\sim m_\pi$ in ordinary nuclei, then it is the standard chiral $n$EFT with the nucleons and pions that is valid. That the standard $n$EFT calculated up to  N$^m$LO for $m\geq 3$ for nuclear dynamics at nuclear matter density $\sim n_0$ works out very well is tantamount to ``proving" the ``FT" for the cutoff scale $\Lambda\sim (400-500)$ MeV. Up to this point Gerry had no disagreement with Weinberg's strategy.
 
In fact  in his 1997 conference talk~\cite{FT} Weinberg acknowledged his agreement with that {\it  what ``they"\footnote{I am supposing Weinberg was referring to Gerry Brown with whom he was in discussions and his colleagues.}  have been doing all along using two-body potentials, including one-pion exchange and a hard core is the correct first step in a consistent approximation scheme.} Furthermore why that works, Weinberg states, is because {\it although nucleons are not soft they never get far from their mass shell, and for that reason can be also treated as the soft pions.} Indeed the prominent role of soft-pion kinematics in nuclear dynamics had been already seen from the ``chiral filter mechanism" observed in the EW response functions obtained before. The great advantage over  what had been done before the arrival of $n$EFT -- which one can say is totally {\it trivial} when viewed from the point of view of the ``FT" -- is that the soft-pion effects are just the leading-order terms in the series one can {\it systematically} write down in power expansion (typically in derivatives and pion mass terms) and verify that the corrections to the soft terms are {\it naturally} suppressed. The corollary to the ``FT" could then be that if higher-order corrections to the given order are not {\it naturally} suppressed, then  something must be amiss in the given EFT Lagrangian treated to that order and the missing elements must be implemented to {\it better  satisfy} the ``FT." Gerry Brown's proposal was to incorporate {\it explicitly} what are considered as relevant massive H degrees of freedom thereby modifying higher derivative terms in the standard $n$EFT that incorporate``hard core" and other non-soft DoFs . For clarity, let me generically call the resulting EFT  ``$Gn$EFT" to be distinguished from the standard $n$EFT.

This is at this point, simultaneously with the birth of $n$EFT, that the notion of hidden local (gauge) theory and hidden scale symmetry first figured in nuclear theory~\cite{BR91}. This paper, often associated with  ``BR scaling" and wrongly interpreted by most of the experimentalists and theorists working on dilepton production in relativistic heavy ion collisions (to be explained below),  extended the scale of $n$EFT to the cutoff relevant to the H mesons, $\Lambda\gsim m_V$, thus bringing in the nucleons $\psi$, vector mesons $V=(\rho, \omega)$ and scalar dilaton $\sigma$.  This $Gn$EFT is to go beyond the regime in which $n$EFT has worked successfully, namely,  the density regime $n\lsim 2 n_0$, to the regime relevant to compact stars, i.e.,  $n_{\rm star}\sim (5-7) n_0$. The HLS fields are taken to be dynamically generated~\cite{HY:PR}   ``composite gauge fields"~\cite{suzuki}.  Thus given that HLS is gauge-equivalent to non-linear sigma model,  pions couple derivatively as in the standard $n$EFT. In addition, the H fields appear at tree order in $Gn$EFT and hence modify the chiral counting rule accordingly. 
\footnote{Initial attempts at the appropriate counting rules were made in \cite{HY:PR} for HLS without fermions and dilaton and in \cite{LMR} for HLS including both fermions and dilaton.  No detailed work  has been made up to today in working out systematic nuclear ``scale-chiral" perturbation theory with the H fields included. There are  two reasons: First there are unfortunately too many unknown parameters to fix completely at higher orders, so awkward, at least naively, to work with, second there have been alternative schemes that have worked for Gerry, e.g.,  the $V_{lowk}$ Wilsonian RG scheme developed by Gerry and his colleagues at Stony Brook. Unless otherwise stated, I will mean this $V_{lowl}$RG framework when referred to as high orders. The high-order scale-chiral PT approach extending the standard chiral perturbation expansion would be welcome and  remains a future problem, hopefully, to be taken up by some of the young generation Gerry had trained who are currently leading experts on $n$EFT.}

There are several profound consequences of the H fields so incorporated~\cite{BR91}\footnote{The problem was addressed in terms of the skyrmion model but the physics is equivalent to the $Gn$EFT in the sense of the Cheshire Cat Principle, deemed consistent in spirit with the premise of the ``FT."}. 

The first notable observation that simplifies the theory is that as the vacuum changes with increasing density $n$ or/and temperature T (that I will call ``background" for short)  the hadron (both the nucleon and the H except for the pion) mass parameters in the $Gn$EFT Lagrangian scale by the condensate ratio $\Phi=\la\sigma\ra^\ast/\la\sigma\ra_0\approx \la\bar{q}q\ra^\ast/\la\bar{q}q\ra_0$ or put in physical terms by the decay constant ratio  $\Phi=f_\sigma^\ast/f_\sigma\approx f_\pi^\ast/f_\pi$ where $\ast$ denotes the background dependence and the subscript $0$ stands for the vacuum.   Such background-dependent effects could partly be included in the phenomenologically fixed constants at higher chiral orders in $n$EFT, but here they are {\it intrinsically} present at the tree order in $Gn$EFT. 

What Gerry was very interested in at that time after the BR scaling was put forward was the {\it effective} $V$ mass at high T that were thought to be measurable in relativistic heavy ion collisions. If the vector meson mass were to scale with $\Phi$, then one would naively expect that as $f_\pi^\ast$  goes to zero as predicted by QCD at high temperature, the $V$ mass would go to zero. This should be visible, it was argued, in dilepton-production processes in relativistic heavy-ion collisions as T approached the chiral transition temperature $T_c$. The proposal based on this observation first made  by Gerry with his co-workers at Stony Brook triggered massive dilepton searches in heavy-ion experiments.   It turned out that  the extensive searches found no signal for such behavior~\cite{braun-munzinger}. The observed spectrum was found to be consistent with what would more or less correspond to mundane hadronic interactions, showing no enhanced dilepton production below the vacuum $\rho/\omega$ mass. This led to the conclusion that dropping mass scenario {\it predicted} by the Brown-Rho scaling  was {\it ruled out}. 

This conclusion was  incorrect due to the basically wrong interpretation of the mass scaling. The mass scaling given in \cite{BR91} was to hold for {\it low T} and {\it low $n$}  but  invalid near phase transitions.  Moreover both Gerry and I quickly realized when the data started to appear that the dilepton measurements  made up to then,  technically superb though they seemed to be,  had very little --  if not nothing --  to do with the dropping $\rho$ mass effect they were looking for. In the HLS-based theory as formulated~\cite{BR91}, such a dropping mass effect, though in principle could not be excluded as mentioned below,  would be virtually impossible to be sorted out in the measurements made so far.

The reason for this conclusion is very simple. In the hidden local symmetry (HLS) framework  adopted in \cite{BR91}, with the vector mesons taken to be dynamically generated composite gauge fields, it is not the pion decay constant $f_\pi^\ast$ going to zero that the background (T or $n$) drives the vector mass to zero, but what is in action is the RG flow of the gauge coupling $g_V$  to  ``the vector manifestation (VM)" fixed point~\cite{HY:PR} at which $g_V\to 0$ whatever $f_\pi^\ast$ may be.  This $g_V\to 0$ makes both the mass and width simultaneously go to zero. Furthermore the $\gamma\pi\pi$ coupling deviates maximally from the vector dominance assumed in {\it all} analyses of the dilepton spectrum. As a consequence, the smoking-gun signal for such $V$ with dropping mass, say, $V^\ast$, if present,   would be a {\it tiny} sharp peak, like a needle buried in a huge haystack~\cite{haystack}!

This was quite a bit disappointing -- although expected by some of us at the time -- particularly for the tour-de-force experimental effort, but it turns out fortunately that the ``failed search" brought out a drastically novel and much more interesting scenario, perhaps a deep issue in nuclear physics,  buried in the haystack. Gerry discussed some of his ideas in \cite{haystack}.  I will come back to this matter below.

The second consequence of the H fields of \cite{BR91} has to do with the long-standing puzzle of ``quenched $g_A$" in nuclear Gamow-Teller transitions. It not only resolves the puzzle and show how the ``FT" works in nuclear physics but also poses a challenge for  a  genuine first-principles nuclear physics calculation.

When the Lagrangian in \cite{BR91} was treated in nuclear matter in the mean-field approximation, one could calculate extremely simply the effective axial-vector coupling constant denoted $g_A^\ast$ in nuclear superallowed Gamow-Teller transitions at near nuclear matter density. The approximation was valid in the large $N_c$ and large $k_F/\Delta E$ (where $\Delta E$ is roughly the excitation energy of the virtual state the GT operator excites) limits. The first has to do with that $g_A\sim O(N_c)$ in QCD and the second with the Fermi-liquid structure of nuclear matter. Being roughly equivalent to what's known as the Landau Fermi-liquid fixed-point (FLFP) approximation, it corresponds to a soft-pion and soft-dilaton approximation.  It is tantamount to decimating in the generating functional the (Wilsonian) path integral from the cutoff above the Fermi surface to the top of the Fermi surface in the FLFP approximation in nuclear medium.  For the superallowed transition, this $g_A^\ast$ multiplying the quasi-nucleon making the GT beta-decay transition  at zero momentum transfer via the operator $\tau^{\pm}\sigma$ on the Fermi surface captures the {\it complete} nuclear correlations coupled to the single-particle Gamow-Teller operator. The resulting $g_A^\ast$ predicted  in \cite{BR91} was surprisingly close to 1, $g^L_A \approx 1$ (the superscript $L$ standing for Landau Fermi liquid)  nealy independently of $k_F$  near nuclear matter density $n_0$.  Put in terms of  what is called in the literature as ``quenching factor" $q$,  it was $q^L\equiv g_A^L/g_A \approx 0.78$ with $g_A=1.276$ from neutron beta decay. This can be applied to light as well as heavy nuclei.

This result $g_A^L\approx 1$ can be connected to what is obtained in nuclear Gamow-Teller transitions in simple shell model. The most unambiguous way is resort to what is known as ``extreme single-particle shell model (ESPSM)." The doubly closed shell nucleus $^{100}$Sn is an example.\footnote{Unfortunately the available experiments are not in agreement with each other. I will return to this matter below.}  It is found in light nuclei in simple shell model that $q^{\rm light}\approx 0.75(2)$. The optimal condition for ESPSM is not strictly met in all light nuclei studied to make a precise comparison with $g_A^L$ but it is not accidental that $q^{\rm light}$ is $g_A^{\rm light}\approx 1$.


As indicated, this problem of the quenched $g_A$ points out how the ``FT" could be intricately involved in the interplay of various nuclear phenomena. In \cite{first-principles}, an attempt is made to resolve the problem from what the authors called ``first principles." They address it by, I quote,  ``a state-of-the-art computations of $\beta$ decays from light- and medium-mass nuclei to $^{100}$ Sn by combining effective field theories of the strong and weak forces with powerful quantum many-body techniques."  In the framework of Weinberg's counting in $n$EFT, what correspond to the the irreducible terms, i.e., the nuclear forces and currents,  are treated respectively to N$^4$LO and to N$^3$LO and the reducible terms are handled in the ``no-core-shell-model," considered as ``virtually exact treatment of correlations in the nuclear wave functions." There is inevitably the issue of (in)consistency in the chiral counting in this sort of approaches.  There is in principle a fully consistent scheme~\cite{KSW},  but it is strictly valid for, e.g.,  pionless EFT with the pions integrated out. And I am not aware of any truly rigorous counting scheme being used in such ``first principles" approach to many-body systems like $^{100}$Sn. 

It was known since a ling time, thanks to the chiral filter argument based on current algebras,  that many-body contributions to the one-body Gamow-Teller operator,  in contrast to the magnetic moments and axial-charge transitions,  should be suppressed, explaining,  among others, why the pp solar neutrino matrix element had only small corrections from two-body currents. The detailed derivation in $n$EFT first performed in early 1990's  indeed showed that the leading two-body (called exchange-current) corrections to the single-particle Gamow-Teller operator come three orders down at N$^3$LO~\cite{parketal}\footnote{The counting rule in $n$EFT for the nuclear EW currents was initiated by Park in early 1990's (for his PhD thesis) after the appearance of Weinberg's papers in which he re-derived  in chiral perturbation theory, among others,  the chiral filter mechanism.  In this development, he was inspired particularly by Gerry's ideas that were shared by other Koreans working in that field. Its publication was unfortunately delayed by  (especially PRL) referees until 2003 when his calculations were further supported thanks to the help of  the Italian colleagues. By now there are quite a few excellent review papers written on the subject improving on Park's derivation.}. Thus it followed from the premise of the ``FT" that leading corrections from many-body currents, if they were to make sense, should be {\it strongly} suppressed in superallowed Gamow-Teller transitions, particularly for the doubly-closed-shell nucleus like $^{100}$Sn. It was found however in the calculation~\cite{first-principles} that the two-body contribution  $q_{\rm 2BC}$ falls in the range compatible with the experimental $q_{\rm ESPSM}$ making the two-body correction to the LO Gamow-Teller operator be $\gsim  25\%$. This is much too big to be considered  natural and consistent with the premise of the  ``FT."\footnote{The data on $^{100}$Sn used in \cite{first-principles} has been improved  on in a more recent measurement that would require even greater two-body contributions. Given the significance of the trace anomaly issue and also for the shell-model description, further experimental works on this nucleus are highly needed.} Apart from the fact that there are more N$^{3}$LO graphs from nonrelativistic corrections such as recoil terms that are not taken into account but cannot be ignored, there are no reasons to stop at the N$^{3}$LO with such big contributions. The next N$^{4}$LO with a huge number of undeterminable  parameters  could equally be non-negligible and so on.  This means that terms of order N$^m$LO with $m\geq 4$ could be totally out of control.  The ``FT" would decree this  cannot be the correct way of doing physics. Something must be amiss in this way of doing EFT. 

A possible solution to the problem~\cite{MRgA} was argued to be in the role played by the hidden H~\cite{BR91}. Since the external axial-vector field couples scale-invariantly to the nucleon with the nuclear EFT defined at the chiral scale,  $\sim 4\pi f_\pi\sim 1$ GeV, the coupling constant $g_A$ does not scale in density.  It is therefore the {\it full nuclear correlation} in the nuclear matrix element of the axial current $\bar\psi\tau^{\pm}\gamma_\mu\gamma_5\psi$ that must account for the quenching factor such that $g_A^\ast \to 1$. 
In \cite{BR91,MRgA}, it is the interplay of the hidden local symmetry and hidden scale symmetry ``emerging" through strong nuclear correlations in nuclear matter encapsulated in the soft theorems that make the N$^m$LO  $m >2$ many-body GT current contribution tend to nearly zero. This tendency is in fact visible in the calculation of \cite{first-principles} when the ``resolution scale" is increased. It would bring the multi-body corrections to near zero if the  resolution scale is made high enough. Doing the {\it full} correlation calculation in the doubly-closed-shell nucleus $^{100}$Sn which could involve an excitation energy of a few hundred MeV connected by the nuclear tensor force is nearly impossible, but in few-body systems,  one should be able to check this argument with the presently available powerful quantum many-body techniques. Indeed  the current state-of-the-art quantum Monte Carlo calculations with N$^{3}$LO axial currents in light nuclei $A < 10$  confirm this assertion at the 2-3\% level without invoking the quenched $g_A$~\cite{wiringa}. This of course does not mean that many-body terms should be simply set equal to zero. It is just that they should be {\it naturally}  suppressed by the chiral-counting. It means in the spirit of the ``FT" that there can be other corrections of the same order coming from sources that go beyond the given EFT scheme. An example is the possible effect of the quantum (trace) anomaly in QCD that breaks scale symmetry explicitly in the axial current coupling to the nucleon involving the anomalous dimension $\beta^\prime$ of the gluon stress tensor. This matter concerns the case of the ESPSM with $^{100}$Sn where there is a disagreement between the available experiments. This may involve a fundamental issue in gauge theories even relevant to the BSM (beyond the standard model) as discussed in \cite{RM2021,MRgA}. 

It is made absolutely clear by now that the original idea of BR scaling~\cite{BR91}  had been neither ``ruled out" \footnote{This dilepton fiasco misled not just (more or less all) heavy-ion physicists but also a large number of nuclear physicists not working in the field of heavy-ion physics.  Just to repeat what has been pointed out before, what's formulated in terms of scale-invariant HLS~\cite{BR91} is,  when done correctly,  essentially equivalent to the standard $n$EFT for the density regime $n\lsim 3n_0$. This applies not only to infinite matter but also to finite nuclei.  An apt case where things work out very well near $n_0$ is the famous C-14 archaeological dating $\beta$ decay process that Gerry and his co-workers explained beautifully using the BR scaling relation~\cite{c-14}. This process has also been explained by what's heralded as ``first-principles {\it ab initio} EFT calculation" where no BR scaling is invoked, which led some people to question the validity of  Gerry et al's explanation based on the scaling relation. But the point is that they were doing the same physics. What's misunderstood is that the scaling Gerry et al used contains in addition to the IDD (``intrinsic density dependence" inherited from QCD) but also the induced contribution coming from 3-body short-range forces mediated by the $\omega$ exchange that are {\it integrated into} the two-body tensor forces crucially intervening in the calculation. This is just what the ``FT" required of the $Gn$EFT. What happens in the higher density regime $n\gsim  3n_0$ relevant to compact-star physics not spelled out in \cite{BR91} is a subtle matter involving topology~\cite{MR-PPNP} which is undergoing further development as mentioned below.}  nor even probed by the dilepton experiments at high temperature. The arrival of holographic dual QCD where the infinite tower of gauge fields figure made this even clearer~\cite{haystack}. There is also the possibility of an infrared fixed point with scale invariance at high temperature, so far unaccounted for by heavy-ion theorists~\cite{IRfp}.  The possible ``duality" of the vector mesons to the gluons of QCD at temperature near $T_c$~\cite{Y}  would mean that the phase transition purported to be probed by dileptons could involve a possible topological phase, totally different from what so far went into the analysis of the process involved. Thus it's a whole new ball game at high temperature. 


Highly dense compact-star matter Gerry became extremely enthusiastic about after the needle-in-the-haystack fiasco illustrates how his line of thought went in the spirit of the ``FT."  Here again the role of the H particles, the ancient idea of  \cite{BR91}, is found to play a potentially crucial role.  The development, which started when Gerry visited in early 2000 the newly established Korean institute called KIAS (Korea Institute for Advanced Study) -- where I was a member -- and has continued in Korea, while Gerry focused on supernova explosions in Stony Brook, is currently making a possible paradigm change in compact-star physics with the observations of gravity waves. 

Gerry was not directly involved in this, but let me recount what amazingly transpired after Gerry's visit to KIAS. The Korean Government established first the World Class University Project followed by the Institute for Basic Science, an ambitious project for fundamental research in Korea.  Gerry was influential in the first project, again indirectly but crucially conceptually.  One of the principal outcomes of the project was to arrive at the equation of state (EoS) of the matter at the highest  baryonic densities in the Universe relevant to massive compact stars. The idea was to again exploit the two hidden symmetries and the Cheshire Cat Principle to formulate a bottom-up approach via $Gn$EFT and establish the ``hadron-quark duality." A surprising result was that while the $Gn$EFT post-dicted, all satisfactorily,  the baryonic matter at  $\sim n_0$ as well as at densities relevant to compact stars, it made a specific prediction that the core of the massive stars is a matter of  ``quasiparticles" of fractional baryonic charges -- behaving neither like baryons nor like deconfined quarks  -- and more surprisingly with the star velocity close to the ``conformal speed" $v_s/c^2\approx 1/3$ although the trace of the energy-momentum tensor is not equal to zero with the non-zero dilaton mass.\footnote{An amusing possibility is that with PC symmetry, the core of massive stars  could be of an  ``unmatter" in analogy to ``unparticle"~\cite{unparticle} and ``unnucleus"~\cite{unnucleus}.}  This prediction is drastically different from others available in the literature. It was interpreted as  the emergence of the hidden scale symmetry or what may be called ``pseudo-conformal symmetry (PCS)" as observed in the case of the quenched $g_A$. This description has no phase transitions.  One could say this is a bottom-up description of the putative hadron-quark continuity in the spirit of the ``FT." There could also be a top-down approach in QCD starting with quarks at high density, going smoothly over to baryons at some density carrying the same star speed  as in the bottom-up description.   It looks very likely that such a top-down approach can be made, by fine-tuning parameters, to signal the same ``emergent" scale symmetry as the bottom-up one. This would be a great progress toward the understanding of an extremely challenging field. For initial efforts, see  \cite{RM2021,MR-PPNP}.  

I will not be detailed in what follows because while closely related to what I described above, it is very little understood, so what I say can be at best speculative. It concerns  surprising ``new" developments  coming from outside of nuclear physics circle on the ``old" vector mesons, which I am certain Gerry would have enthusiastically endorsed as he did with the old idea of HLS. Whatever the case may be, it is definitely novel and intriguing involving some of Gerry's favored ideas, a few of which were expressed in \cite{haystack}.

It has closely to do with something mystifying in the notion of  HLS for the vector mesons $V$ -- and also  of scale symmetry for the dilaton $\sigma$ -- in nuclear dynamics.  Let me limit to the first here.

When HLS Lagrangian is written to the leading order in the power counting, say, $O(p^2)$, the vector meson mass at the tree oder is given by $m_V^2=a f_\pi^2 g_V^2$ where $a$ is an unknown constant parameter in HLS theory. There are several surprises here. The first surprise is that this formula turns out to be valid to all orders of loop corrections~\cite{HY:PR}. Furthermore, though it is a conjecture not  yet rigorously proven,  the HLS fields are considered to be Seiberg-dual to the gluons of QCD~\cite{Y,komargodski,abel}. The duality gives the second surprise,  $a=2$,  leading to the old Sakurai's vector dominance (VD).   Thus $\gamma\pi\pi$ is vector-dominated in the vacuum but gets modified at high temperature or density.  A renormalization group analysis shows that $g_V$ flows to zero and $a$ to 1, responsible for  the needle in the haystack in the dilepton production.

Another surprise  is  in the skyrmion description in setting up the HLS-based formulation in \cite{BR91}. We had in mind the Cheshire Cat Principle~\cite{CCP}  in making contact with both ``soft" and ``hard" degrees of freedom.  There, roughly speaking, what happens is that the pions effectively turn via topology into the nucleons made up of u and d quarks. (Similarly for $N_f=3$, the  octet baryons made up of u, d and s quarks come from the octet mesons.)  What was involved was the ``infinite hotel mechanism (IHM)"~\cite{IHM}.  The flavor singlet $\eta^\prime$, it turns out however,  cannot make the usual Cheshire-Cat transformation via the IHM to a flavor singlet baryon. This is because it is  forbidden by topology. This means that there is no skyrmion coming from $\eta^\prime$. In a rather intricate mechanism, it has been found that the $\eta^\prime$ does actually transform to a topological baryon but the baryon is not a skyrmion but a fractional quantum Hall (FQH) droplet. It is  a two spatial-dimensional pancake-type object,  a domain-wall configuration of the $\eta^\prime$ meson bounded by a string ($\eta^\prime$ singularity or ring) embedded in 3 dimensions~\cite{komargodski-FQH}.  One can in fact concoct a {\it different}  Cheshire Cat mechanism~\cite{MNRZ} which goes not via the IHM but via what is known as ``anomaly inflow mechanism"~\cite{anomaly-inflow} into a sheet structure.  It has the same topological structure of the famous FQH effect encoded in Chern-Simons gauge fields in condensed matter physics. This raises the dichotomy problem between the two topological objects, the skyrmion and the FQH pancake,  because a baryon of large spin -- for large $N_c$ -- such as $\Delta (3/2,3/2)$ can be described both as a skyrmion and as a FQH pancake~\cite{komargodski-FQH}.  But in what way they are connected is not known, hence the ``dichotomy."  The new development is that this problem {\it must} be resolved to understand from the point of view of the ``FT"  what actually happens at large density or large temperature. Yet another surprise! The resolution, it is argued, must -- again -- involve hidden local symmetry.

Such a baryon  coming from $\eta^\prime$, if it exists,  must be a metastable object since it is not seen in the vacuum. So at normal density or low temperature, it could very well be irrelevant in nuclear physics.  However approaching phase transitions such as chiral transition and/or deconfinement, the current  argument goes, it must become not just relevant but even indispensable. It is not understood what that means for experimental observables at high density or temperature  but there are intriguing possibilities that cannot be ruled out by the currently available theoretical tools.  

Among the various ideas Gerry put forward~\cite{haystack}, one possibility was the  ``hadronic freedom"  -- and associated quantities -- at near $T_c$ that has remained unexplored by heavy-ion theorists.  As far as I am aware, it is neither confirmed nor refuted.  It concerns an uncharted domain, both in temperature and in density.  

So the question Gerry left behind was: What can it be?

At high temperature with the dimensional reduction,  the vector mesons $V$ could be taken, thanks to Seiberg-type dualities,  as the gauge bosons of the $U(N_f)_{N_c}$  Chern-Simons theory  ``dual" to the gluons of QCD~\cite{Y}.  Now the  $\eta^\prime$ meson bounded by a string, i.e., the  FQH pancake~\cite{komargodski-FQH}, while invisible at low T,   it has been recently argued~\cite{vectors-on-wall,karasik}, becomes {\it indispensable} for describing the vector mesons in hidden local symmetry when they are near the chiral phase transition. What happens is that approaching the transition, the vector mesons enable the FQH pancakes to appear and transform to  the Chern-Simons topological gauge fields~\cite{vectors-on-wall}. The properties, such as  coupling to photons, of such vector mesons in the topological phase -- instead of in the Higgs phase -- could then be totally different from the standard treatment of high-T matter in heavy-ion collisions, e.g.,  in analyzing the dilepton data~\cite{braun-munzinger}.  

They could also play an intriguing role at high density in compact stars where the hidden symmetries dictate the structure of the high-density core of massive stars~\cite{RM2021}.    

These ``exotic" phenomena may ultimately be accessible top-down from QCD proper.  At present, however, they remain totally unexplored even in temperature for which lattice QCD is available.  Gerry had various ideas to proceed bottom-up via $Gn$EFT in the spirit of the ``FT," but he left them to the future generations.


\begin{thebibliography}{50}
\bibitem{wilkinson}  {\it Progress in Particle and Nuclear Physics: Quarks and the Nucleus} \ (Pergamon Press, Oxford, 1982) ed. Denys Wilkinson 
\bi{schwinger-fest} S.~Weinberg,
  ``Phenomenological Lagrangians,''
  Physica A {\bf 96}, no. 1-2, 327 (1979).
      
\bi{CCP}   S.~Nadkarni, H.~B.~Nielsen and I.~Zahed,
  ``Bosonization relations as bag boundary conditions,''
  Nucl.\ Phys.\ B {\bf 253}, 308 (1985);  S.~Nadkarni and I.~Zahed,
  ``Nonabelian Cheshire Cat bag models in (1+1)-dimensions,''
  Nucl.\ Phys.\ B {\bf 263}, 23 (1986);  P.~H.~Damgaard, H.~B.~Nielsen and R.~Sollacher,
  ``Smooth bosonization: The Cheshire cat revisited,''  Nucl.\ Phys.\ B {\bf 385}, 227 (1992).  
  
 \bi{weinberg1} S.~Weinberg,
``Nuclear forces from chiral Lagrangians,'' Phys.\ Lett.\ B {\bf 251}, 288 (1990).

\bi{weinberg2} S.~Weinberg,  ``Effective chiral Lagrangians for nucleon - pion interactions and nuclear forces,'' Nucl.\ Phys.\ B {\bf 363}, 3 (1991).


\bi{MR91} M.~Rho,
  ``Exchange currents from chiral Lagrangians,''
  Phys.\ Rev.\ Lett.\  {\bf 66}, 1275 (1991).  
  
  
   
\bi{FT} S. Weinberg, 
  ``What is quantum field theory, and what did we think it is?,''
  In *Boston 1996, Conceptual foundations of quantum field theory* 241-251
  [hep-th/9702027].
  
  
  \bi{BR91} G.~E.~Brown and M.~Rho,
  ``Scaling effective Lagrangians in a dense medium,''
  Phys.\ Rev.\ Lett.\  {\bf 66}, 2720 (1991). 

  
 \bi{pionless} H.-W.~Hammer, S.~K\"onig and U.~van Kolck,
  ``Nuclear effective field theory: status and perspectives,''
  Rev.\ Mod.\ Phys.\  {\bf 92}, no. 2, 025004 (2020).  
  
     
  \bi{HY:PR} M.~Harada and K.~Yamawaki,
  ``Hidden local symmetry at loop: A New perspective of composite gauge boson and chiral phase transition,''
  Phys.\ Rept.\  {\bf 381}, 1 (2003).
  
  \bi{suzuki}   M.~Suzuki, ``Inevitable emergence of composite gauge bosons,''
  Phys.\ Rev.\ D {\bf 96}, no. 6, 065010 (2017) .
  
  \bi{LMR} Y.~L.~Li, Y.~L.~Ma and M.~Rho,
  ``Chiral-scale effective theory including a dilatonic meson,''
  Phys.\ Rev.\ D {\bf 95}, no. 11, 114011 (2017).
  
  \bi{braun-munzinger}  P.~Braun-Munzinger, V.~Koch, T.~Sch\"afer and J.~Stachel,
  ``Properties of hot and dense matter from relativistic heavy ion collisions,''
  Phys.\ Rept.\  {\bf 621}, 76 (2016).  
  
 \bi{haystack} G.~E.~Brown, M.~Harada, J.~W.~Holt, M.~Rho and C.~Sasaki,
  ``Hidden local field theory and dileptons in relativistic heavy ion collisions,''
  Prog.\ Theor.\ Phys.\  {\bf 121}, 1209 (2009).
  
   
\bi{RM2021} M.~Rho and Y.~L.~Ma,
  ``Manifestation of hidden symmetries in baryonic matter: From finite nuclei to neutron stars,'' Mod.  Phys. Lett. A, 2130012 (2021),
  arXiv:2101.07121 [nucl-th].
  
       
  \bi{first-principles}  P.~Gysbers {\it et al.},
  ``Discrepancy between experimental and theoretical $\beta$-decay rates resolved from first principles,''
  Nature Phys.\  {\bf 15}, no. 5, 428 (2019).
  
 \bi{KSW}   D.~B.~Kaplan, M.~J.~Savage and M.~B.~Wise,
  ``A New expansion for nucleon-nucleon interactions,''
  Phys.\ Lett.\ B {\bf 424}, 390 (1998).

  
 \bi{parketal} 
  T.~S.~Park {\it et al.},
  ``Parameter free effective field theory calculation for the solar proton fusion and hep processes,''
  Phys.\ Rev.\ C {\bf 67}, 055206 (2003) . 
  
 \bi{MRgA} Y.~L.~Ma and M.~Rho,
  ``Quenched $g_A$ in nuclei and emergent scale symmetry in baryonic matter,''
  Phys.\ Rev.\ Lett.\  {\bf 125}, no. 14, 142501 (2020).
  
 \bi{wiringa} G.~B.~King, L.~Andreoli, S.~Pastore, M.~Piarulli, R.~Schiavilla, R.~B.~Wiringa, J.~Carlson and S.~Gandolfi,
  ``Chiral effective field theory calculations of weak transitions in light nuclei,''  Phys.\ Rev.\ C {\bf 102}, no. 2, 025501 (2020).   
  
 \bi{c-14} J.~W.~Holt, G.~E.~Brown, T.~T.~S.~Kuo, J.~D.~Holt and R.~Machleidt,
  ``Shell model description of the C-14 dating beta decay with Brown-Rho-scaled NN interactions,''
  Phys.\ Rev.\ Lett.\  {\bf 100}, 062501 (2008).
    
  \bi{MR-PPNP} Y.~L.~Ma and M.~Rho,
  ``Towards the hadron-quark continuity via a topology change in compact stars,''
  Prog.\ Part.\ Nucl.\ Phys.\  {\bf 113}, 103791 (2020).  
  
 \bi{IRfp} A.~Alexandru and I.~Horv\'ath,
  ``Possible new phase of thermal QCD,'' Phys.\ Rev.\ D {\bf 100}, no. 9, 094507 (2019)
  [arXiv:1906.08047 [hep-lat]]. 
  
   
    \bi{Y} N.~Kan, R.~Kitano, S.~Yankielowicz and R.~Yokokura,
  ``From 3d dualities to hadron physics,''
  Phys.\ Rev.\ D {\bf 102}, no. 12, 125034 (2020).  
  
  \bi{unparticle} H.~Georgi,
  ``Unparticle physics,''
  Phys.\ Rev.\ Lett.\  {\bf 98}, 221601 (2007) .
   
  \bi{unnucleus} 
  H.~W.~Hammer and D.~T.~Son,
  ``Unnuclear physics,'' arXiv:2103.12610 [nucl-th].  
  
 \bi{komargodski} Z.~Komargodski,
  ``Vector Mesons and an interpretation of Seiberg duality,''
  JHEP {\bf 1102}, 019 (2011). 
  
 \bi{abel}  S.~Abel and J.~Barnard,
  ``Seiberg duality versus hidden local symmetry,''
  JHEP {\bf 1205}, 044 (2012).
  
 \bi{IHM}  
  H.~B.~Nielsen and A.~Wirzba,
  ``The Cheshire Cat Principle applied to hybrid bag models,''
  NBI-HE-87-32.  
    
    
 \bi{komargodski-FQH} Z.~Komargodski,
  ``Baryons as quantum Hall droplets,''
  arXiv:1812.09253 [hep-th]. 
  
 \bi{MNRZ} Y.~L.~Ma, M.~A.~Nowak, M.~Rho and I.~Zahed,
  ``Baryon as a quantum Hall droplet and the Cheshire Cat Principle,''
  Phys.\ Rev.\ Lett.\  {\bf 123}, 172301 (2019).
  
 \bi{anomaly-inflow}   C.~G.~Callan, Jr. and J.~A.~Harvey,
  ``Anomalies and fermion zero modes on strings and domain Walls,''
  Nucl.\ Phys.\ B {\bf 250}, 427 (1985).  
  
 \bi{vectors-on-wall} R.~Kitano and R.~Matsudo,
  ``Vector mesons on the wall,'' JHEP {\bf 2103}, 023 (2021).
  
 \bi{karasik}  A.~Karasik,
  ``Vector dominance, one flavored baryons, and QCD domain walls from the ``hidden" Wess-Zumino term,''
  arXiv:2010.10544 [hep-th]; ``Skyrmions, quantum Hall droplets, and one current to rule them all,''
  SciPost Phys.\  {\bf 9}, 008 (2020).
    
\end{thebibliography}
\end{document}